\documentclass{llncs}

\font\twelverm=cmr12
\font\twelveit=cmti12
\font\twelvebf=cmbx12

\font\twelvesl=cmsl12
\font\twelvett=cmtt12

\font\ninebf=cmbx9

\font\sevenrm=cmr7

\font\sevenbf=cmbx7
\font\seveni=cmmi7
\font\sevensy=cmsy7

\font\scshapex=cmcsc10
\font\scshapexii=cmcsc12
\font\scshapeeight=cmcsc10 at 8pt
\font\scshapevii=cmcsc10 at 7pt
\font\scshapev=cmcsc10 at 5pt
\def\scshape{\scshapex}

\newfam\scfam
\textfont\scfam\scshapex
\scriptfont\scfam\scshapevii
\scriptscriptfont\scfam\scshapev

\font\teneufm=eufm10
\font\seveneufm=eufm7
\font\fiveeufm=eufm5

\newfam\eufmfam
\textfont\eufmfam=\teneufm
\scriptfont\eufmfam=\seveneufm
\scriptscriptfont\eufmfam=\fiveeufm

\font\tenmsbm=msbm10
\font\sevenmsbm=msbm10 at 7pt
\font\fivemsbm=msbm10 at 5pt

\newfam\msbmfam
\textfont\msbmfam=\tenmsbm
\scriptfont\msbmfam=\sevenmsbm
\scriptscriptfont\msbmfam=\fivemsbm

\def\mathbb{\fam\msbmfam\tenmsbm}

\def\twelvepoint{\def\rm{\fam0\twelverm}
\textfont0=\tenrm \scriptfont0=\sevenrm \scriptscriptfont0=\fiverm
\textfont1=\teni \scriptfont1=\seveni \scriptscriptfont1=\fivei
\textfont2=\tensy \scriptfont2=\sevensy \scriptscriptfont2=\fivesy
\textfont3=\tenex \scriptfont3=\tenex \scriptscriptfont3=\tenex
\textfont\itfam=\twelveit \def\it{\fam\itfam\twelveit}%
\textfont\slfam=\twelvesl \def\sl{\fam\slfam\twelvesl}%
\textfont\ttfam=\twelvett \def\tt{\fam\ttfam\twelvett}%
\textfont\bffam=\twelvebf \scriptfont\bffam=\ninebf
\scriptscriptfont\bffam=\sevenbf \def\bf{\fam\bffam\twelvebf}%
\normalbaselineskip=14pt
\let\sc=\tenrm \let\big=\twelvebig \normalbaselines\rm
\def\scshape{\scshapexii}
}

\newcommand{\R}{{\mathbb R}}

\newcommand{\N}{{\mathbb N}}

\newcommand\cB{{\cal B}}

\newcommand\cD{{\cal D}}
\newcommand\cE{{\cal E}}

\newcommand\cK{{\cal K}}
\newcommand\cL{{\cal L}}

\newcommand\cV{{\cal V}}

\newcommand{\E}[1]{{\mathbb E}\left[#1\right]}

\newcommand{\p}[1]{{\mathbb P}\left\{#1\right\}}

\newcommand{\eql}{\,{\buildrel \cL \over =}\,}

\newcommand{\eqd}{\,{\buildrel {\rm def} \over =}\,}

\newcommand{\seq}[1]{{(#1)_{t \ge 0}}}
\newcommand{\seqone}[1]{{(#1)_{t \ge 1}}}

\newcommand{\square}{\qed}

\newcommand{\tablerule}{\noalign{\smallbreak \hrule \smallbreak}}
\newcommand{\tabletoprule}{\noalign{\hrule \smallbreak}}
\newcommand{\tablebottomrule}{\noalign{\smallbreak \hrule}}

\newcommand{\idspace}{\{0,1\}^d}
\newcommand{\kbuckets}{$k$-buckets}
\newcommand{\kbucket}{$k$-bucket}
\newcommand{\bi}[1]{ {\cB}_i(#1)}
\newcommand{\bix}{\bi{x}}
\newcommand{\di}[1]{ {\cD}_i(#1)}
\newcommand{\dix}{\di{x}}
\newcommand{\dht}{{\scshape dht}}

\newcommand{\id}{{\scshape id}}
\newcommand{\ids}{{\id s}}
\newcommand{\xor}{{\scshape xor}}
\newcommand{\ptwop}{{\scshape p{\scshapeeight 2}p}}
\newcommand{\polar}[1]{\widehat{#1}}
\newcommand{\pT}{\polar{T}}
\newcommand{\pTi}[1]{{T_{X_#1{\polar{X}_#1}}}}
\newcommand{\ck}{{c_k}}
\newcommand{\ckp}{{c_k'}}
\newcommand{\cks}{{c_k^*}}

\newcommand{\origin}{\bar{0}}
\newcommand{\polarorigin}{\bar{1}}
\newcommand{\xone}{{x\polarorigin}}

\newcommand{\Ws}{{\seq{W_t}}}
\newcommand{\Ss}{{\seq{|S_t|}}}
\newcommand{\Bs}{{\seq{B_t}}}

\newcommand{\Vertices}{\cV}
\newcommand{\Edges}{\cE}

\usepackage[utf8]{inputenc}
\usepackage{amsmath}
\usepackage{array}
\usepackage{tikz}
\usepackage{float}
\usepackage[hidelinks]{hyperref}
\usepackage{pstricks}
\usetikzlibrary{backgrounds,calc,fit,decorations.pathreplacing}  
\usetikzlibrary{positioning,shapes,shadows,arrows,decorations.pathmorphing}

\title{A Probabilistic Analysis of Kademlia Networks\thanks{We would like to
    thank Carlton Davis, Jos\'{e} M. Fernandez, (authors of \cite{Davis2008}
    and \cite{Davis2008b}) and Mahshid Yassaei for collaborations and
explanations regarding Kademlia.}}

\author{Xing Shi Cai \and Luc Devroye}

\institute{
    School of Computer Science, McGill University of Montreal, Canada,\\
    \email{xingshi.cai@mail.mcgill.ca} \\
    \email{lucdevroye@gmail.com}
}

\begin{document}

\maketitle

\begin{abstract}
Kademlia~\cite{Maymounkov02} is currently the most widely used searching
algorithm in \ptwop\ (peer-to-peer) networks.  This work studies an essential
question about Kademlia from a mathematical perspective: how long does it take
to locate a node in the network?  To answer it, we introduce a random graph
$\cK$ and study how many steps are needed to locate a given vertex in $\cK$
using Kademlia's algorithm, which we call the {\em routing time}. Two slightly
different versions of $\cK$ are studied. In the first one, vertices of $\cK$
are labeled with fixed \ids. In the second one, vertices are assumed to have
randomly selected \ids. In both cases, we show that the routing time is about
$c \log n$, where $n$ is the number of nodes in the network and $c$ is an
explicitly described constant.
\end{abstract}

\section{Introduction to Kademlia}

\label{sec:intro}

A \ptwop\ (peer-to-peer) network~\cite{Schollmeier2001} is a computer network
which allows sharing of resources like storage, bandwidth and computing power.
Unlike traditional client-server architectures, in \ptwop\ networks, a
participating computer (a {\em node\/}) is not only a consumer but also a
supplier of resources. Nowadays, major \ptwop\ services in the internet often have
millions of users. For an overview of \ptwop\ networks, see
Steinmetz~\cite{Steinmetz2005}.

The huge size of \ptwop\ networks raises one challenge---among millions of
nodes, how can a node find another one efficiently?  To address this, a group
of algorithms called \dht\ (Distributed Hash Table)~\cite{Balakrishnan03} was
invented in the early 2000s, including Pastry~\cite{Rowstron01}, {\scshape
can}~\cite{Ratnasamy2001}, Chord~\cite{Stoica2001}, Tapestry~\cite{Zhao04}, and
Kademlia~\cite{Maymounkov02}.  Created by Maymounkov and Mazi\`eres in 2002,
Kademlia has become the de facto standard searching algorithm for \ptwop\
services.  It is used by BitTorrent~\cite{Crosby07} and the Kad
network~\cite{Steiner07}, both of which have more than a million nodes.

Kademlia assigns each node an \id\ chosen uniformly at random from $\idspace$, the
$d$-dimension hypercube, where $d$ is usually $128$~\cite{Steiner07} or
$160$~\cite{Crosby07}. Thus we {\em always} refer to a node by its \id. Given two \ids\
$x=(x_1,\ldots,x_d)$ and $y=(y_1,\ldots,y_d)$, Kademlia defines their \xor\
{\em distance} by
$$
\delta(x,y) = \sum_{i=1}^d (x_i \oplus y_i) \times 2^{d-i},
$$
where $\oplus$ denotes the \xor\ operation
$$
u \oplus v =
\begin{cases}
    1  & \hbox{ if $ u \ne v $,} \cr
    0 &  \hbox{ otherwise.} \cr
\end{cases}
$$
Note that {distance} and {closeness} {\em always} mean \xor\
{distances} between \ids\ in this work.

Since it is almost impossible for a node to know where all other nodes are
located, in Kademlia a node, say $x$, is only responsible for
maintaining a table (a {\em routing table\/}) for a small number of other nodes
($x$'s {\em neighbors\/}). Roughly speaking, $\idspace$ is partitioned into $d$
subsets, such that nodes in the same subset have similar distances to $x$.
Within each subset, up to $k$ nodes' information is recorded in a list (a {\it
\kbucket\/}), where $k$ is a constant which usually equals $8$~\cite{Crosby07},
$10$~\cite{Steiner07} or $20$~\cite{Falkner2007}.  All of $x$'s
\kbuckets\ form $x$'s routing table.

When $x$ needs to locate node $y$ which is not in its routing table, $x$ sends
queries to $\alpha$ of its neighbors which are closest to $y$, where $\alpha$
is a constant, sometimes chosen as $3$~\cite{Steiner2007} or
$10$~\cite{Falkner2007}. A recipient of $x$'s message returns locations of $k$
of its own neighbors with shortest distance to $y$.  With this information, $x$
again contacts $\alpha$ nodes that are closest to $y$. This approach ({\it
routing}) repeats until no one closer to $y$ can be found.  The efficiency of
routing is critical for the overall performance of Kademlia.  Its analysis is
the topic of this work.

\section{Our model}

\label{sec:model}

\begin{figure}
\centering {
    \scalebox{0.8} {
        \includegraphics{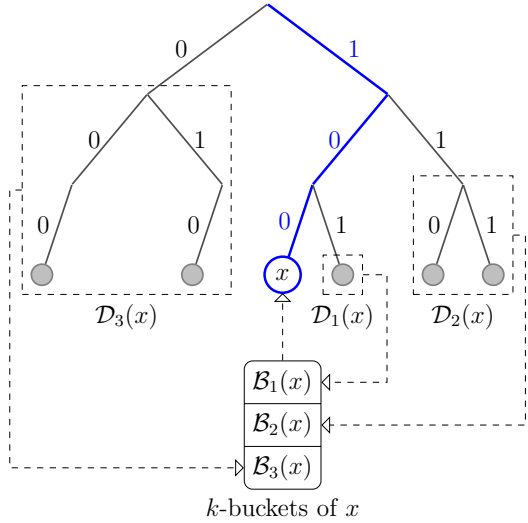}
    }
}
\caption[]{An example of Kademlia \id\ trie and $k$-buckets. Given
an \id\ $x=(1,0,0)$, the trie is partitioned into subtrees $\cD_1(x), \cD_2(x),
\cD_3(x)$.  Node $x$ maintains a $k$-bucket for each of these subtrees containing
the information of up to $k$ nodes in the subtree, which we denote by
$\cB_1(x), \cB_2(x), \cB_3(x)$ respectively.}
\label{fig:partition}
\end{figure}

Consider a Kademlia network of $n$ nodes $X_1,\ldots,X_n$. Writing \ids\ as
strings consisting of zeros and ones, from higher bits to lower bits, we can
completely represent $X_1,\ldots,X_n$ in a binary {\em trie}, as depicted in
Fig.\,\ref{fig:partition}. A trie is an ordered tree data structure invented by
Fredkin~\cite{Fredkin1960}. For more on tries, see
Szpankowski~\cite{Szpankowski2011}. Paths are associated with bit strings---$0$
corresponds to a left child, and $1$ to a right child. The bits encountered on
a path of length $d$ to a leaf is the \id, or value, of the leaf. In this
manner, the binary trie, has height $d$, and precisely $n$ leaves of distance
$d$ from the root.

Let $x = (x_1,\ldots,x_d)$ and $y =
(y_1,\ldots,y_d)$ be two \ids\ (leaves) in the trie.  Let
$\ell(x,y)$ be the length of the path from the root to $x$ and $y$'s lowest
common ancestor, i.e., the length of $x$ and $y$'s common prefix. We have
$$
\ell(x,y) = \max\{i:x_1=y_1,\ldots,x_i=y_i\} \enspace.
$$
It is easy to verify that $\ell(x,y)$ bounds the distance of $x$ and $y$ by
$$
2^{d - \ell(x,y) - 1} \le \delta(x,y) < 2^{d - \ell(x,y)}  \enspace.
$$
Thus if we partition $\idspace \setminus \{x\}$ according to
distances to $x$ as follows
$$
\dix = \{y:2^{i-1} \le \delta(x,y) < 2^i\}, \quad i =
    1,\ldots, d,
$$
then ${\cD}_i(x)$ is equivalent to a subtree in the trie, in which each node
shares a common prefix of length $i$ with $x$. Therefore, we have the
equivalent definition
$$
\dix = \{y:\ell(x,y) = d-i\}, \quad i = 1,\ldots, d  \enspace.
$$

Let $\bix$ be the set of \ids\ in the \kbucket\ of $x$ corresponding to $\dix$. In our
model, we assume that if $|\dix| \le k$, then $\bix = \dix$. Otherwise, for all
$A \subset \dix$ with $|A| = k$, we have
$$
\p{\bix = A} 
    = {{|\dix|} \choose k}^{-1}  \enspace.
$$
In other words, we fill up each \kbucket\ uniformly at random without
replacement.

Consider a directed graph $\cK$ with vertex set
${\Vertices} = \{X_1,\ldots,X_n\}$. Let its edge set be
$$
\Edges = \{(u,v):u,v \in \Vertices, \, v \in \cup_{i=1}^d \bi{u}\}  \enspace.
$$
Put differently, we add a directed edge $(u,v)$ in $\cK$ if and only if $v$ is in one
of $u$'s \kbuckets.  When $\alpha = 1$, only one node is queried at each
step of routing, the search process starting at $x$ for $y$ can be seen as
a path $\rho_{xy}$ in $\cK$ (the {\em routing path}). It starts from vertex
$x$, then jumps to the vertex that is closest to $y$ among all $x$'s neighbors.
From there, it again jumps to the adjacent vertex that is closest to $y$. Let
$y^*$ be the unique vertex with the shortest distance to $y$ in $\cK$. 
Since the distance between the current vertex and $y^*$ decreases at each step until
zero, $\rho_{xy}$ has no loop and always ends at $y^*$.  In fact, $\cK$ is
always {strongly connected}.

Let $T_{xy}$ be the path length of $\rho_{xy}$.  Since $T_{xy}$ equals the
number of rounds of messages $x$ needs to send before routing ends, we call it
the {\em routing time}.  Our first main result assumes that
$X_1=x_1,\ldots,X_n=x_n$, where $x_1,\ldots,x_n$ are all fixed $d$-bit vectors,
which we call the {\em deterministic \id\ model}. The randomness thus comes
only from the filling of the \kbuckets. We always assume that $d \ge \log_2 n$,
and we let $n \to \infty$ (and thus $d \to \infty$) in our asymptotic results.
(Note that in this work $\log n$ denotes the natural logarithm of $n$.)
\begin{theorem}
\label{thm:speed:upper}
In the deterministic \id\ model, we have
\begin{align*}
    \sup_{x_1,\ldots,x_n} \sup_{x} \, \sup_{y} \E{T_{xy}} \le (\ck + o(1)) {{\log n}}, \cr
    \sup_{x_1,\ldots,x_n} \sup_{x} \E{\sup_{y} T_{xy}} \le (\ckp + o(1)) {{\log n}}, \cr
    \sup_{x_1,\ldots,x_n} \E{\sup_{x} \, \sup_{y} T_{xy}} \le (\cks + o(1)) {{\log n}},
\end{align*}
where $\ck, \ckp,\cks$ are constants depending only on $k$.
In particular, we have $\ck = 1/H_k$, where $H_k = \sum_{i=1}^k 1/i$,
also known as the $k$-th harmonic number.
\end{theorem}

\noindent The first inequality in this theorem gives an upper bound over the
expected routing time between two fixed nodes. Since $c_k \le c_1 \le 1/\log
2$, the first bound is better than the $\left\lceil \log_2 n \right\rceil$ bound
described by Maymounkov and {Ma\-zi\`e\-res} in the original Kademlia
paper~\cite{Maymounkov02}. The second inequality considers the expectation of
the maximal routing time when the starting node is fixed, and a look-up is
performed for each of the $n$ destinations. The third one considers the
expectation of the maximal routing time in the whole network if all $n$ nodes were
to look up all $n$ destinations.

Our second main result considers the situation when $X_1,\ldots,X_n$ are
selected uniformly at random from $\idspace$ without replacement (the {\em random
\id\ model\/}).  Given an \id\ $x$, let $\polar{x}$ denote the \id\ that is
farthest away from $x$ ($x$'s {\em polar opposite}).
Since by symmetry $\pTi{1}, \pTi{2}, \ldots \pTi{n}$ are
identically distributed, we only need to study $\pTi{1}$, which we denote by
$\pT$.
\begin{theorem}[\cite{Cai2012}]
    \label{thm:distribution:t:polar}
    In the random \id\ model, we have
    $$
    {\pT \over {\log n}} \to {1 \over g(k)} \qquad \hbox{in probability}, 
    $$
    as $n \to \infty$, where $g(k) = H_k + O(1)$ is a function of $k$.
\end{theorem}
Since $\pT = \pTi{1}$, we see that
$$
{1 \over g(k)} \le {1 \over {H_k}}  \enspace.
$$
However, we have almost identical behavior because $g(k)= H_{k} + O(1)$ as $k
\to \infty$. Let $d=d(n) \ge \log_2 n$. By the probabilistic method,
Theorem~\ref{thm:distribution:t:polar} implies that for every $\epsilon > 0$, there
exists for every $n$ a non-repetitive sequence $(x_{1}(n),\ldots,x_{n}(n)) \in
(\idspace)^n$ of
deterministic \ids\ such that with probability going to one,
$$
\pTi{1} \ge \left( {1 \over {g(k)}} - \epsilon \right) \log n  \enspace.
$$
In view of $g(k) = H_k + O(1)$, we thus see that the first bound of
Theorem~\ref{thm:speed:upper} is almost tight.

Recall that $y^*$ denotes the node that is closest to \id\ $y$. If we look at
the lowest common ancestor of $y^*$ and each hop of $\rho_{xy}$, then searching
$y$ from $x$ can be seen as travelling downwards along the path from the root
to the leaf $y^*$ in the trie, with the distance of each hop being random. In
the random \id\ model, when $n$ is large, we can approximate this process in a
full binary trie with infinite depth.  In Sect.\,\ref{sec:random}, we sketch the proof
of Theorem \ref{thm:distribution:t:polar} which uses this method. This does not work in
the deterministic \id\ model as the structure of the \id\ trie can be
unbalanced. But in Sect.\,\ref{sec:deterministic}, we show that there is another way
to bound the routing time.

Due to its success, Kademlia has aroused great interest among researchers. But
this is the first time that it is studied from a mathematical perspective. Our
results point out one important reason for the success of Kademlia --- its
routing algorithm, while being extremely simple, works surprisingly well. This
work also shows that probabilistic methods together with the right choice of a
data structure, a trie in our case, could significantly simplify the analysis
of a problem which was previously considered too troublesome to analyze
rigorously.

\section{The deterministic ID model}
\label{sec:deterministic}

In this section, we assume that $X_1 = x_1,\ldots,X_n = x_n$, where
$x_1,\ldots,x_n$ are fixed $d$-bits vectors.  Note that the distribution of
$T_{xy}$ is decided only by distances between vertices and the distance between
$x$ and $y$.  Thus, by rotating the hypercube, we can always assume
$y$ to be a specific \id, which we choose $\polarorigin = (1,1,\ldots,1)$.

\begin{figure}
\centering {
    \scalebox{0.8} {
        \includegraphics{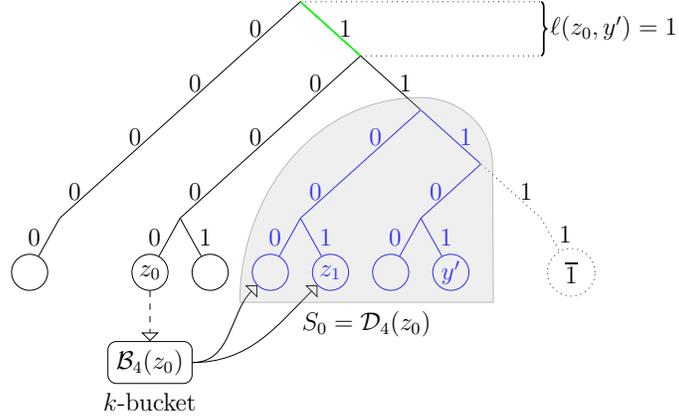}
    }
}
\caption{An example of the first hop of $\rho_\xone$ with $d=5$,
    $k=2$.
    Since $d - \ell(z_0, y') = 4$, $z_1$ must be in subtree $S_0 =
    \cD_4(z_0)$, which is the highest subtree to the right of $z_0$. We
    choose up to $k$ leaves from $S_0$ uniformly at random without
replacement, and let $z_1$ be the rightmost one.}
\label{fig:routing}
\end{figure}

Figure~\ref{fig:routing} depicts the first hop of $\rho_\xone$ as jumping from one
leaf to another in the \id\ trie.  It
is easy to see that if we always arrange branches representing $1$ to the right
hand side, which we take as a convention, then the closer a leaf is to the
right,  the closer it is to $\polarorigin$. Thus the rightmost leaf in the trie,
which we always denote by $y'$, is closest to $\polarorigin$ and is thus
the end point of $\rho_\xone$. 

Write $\rho_\xone = (z_0,z_1,\ldots)$ where $z_0 = x$.  Let $i = d -
\ell(z_0,\polarorigin)$. We can see from Fig.\,\ref{fig:routing} that $z_1$, the
first hop, must belong to $\cD_i(z_0)$, the highest subtree on the right hand
side of $z_0$, which we denote by $S_0$. Since being $z_0$'s neighbor implies
membership in one of $z_0$'s \kbuckets, we have $z_1 \in \cB_i(z_0) \subseteq
S_0$.  Recalling how $\cB_i(z_0)$ is decided, we can think of the first hop as
selecting up to $k$ leaves from $S_0$ uniformly at random and choosing the
rightmost one as $z_1$.  Thus we can define of $\rho_\xone$ recursively as
follows:

\begin{itemize}
    \item Let $z_0 = x$. Repeat the following step.

    \item Given $z_t$ and $t \ge 0$, let $S_t$ be highest subtree on the
    right hand side of $z_t$. If $S_t = \emptyset$, 
    terminate. Otherwise select up to $k$ leaves from $S_t$ uniformly at
    random without replacement, and let the rightmost one be $z_{t+1}$.
\end{itemize}

Since $S_0 \supset S_1 \supset \dots \supset S_{T_\xone}=\emptyset$,  by
studying how quickly the sequence $\Ss$ decreases to $0$, we can bound how big
$T_\xone$ could be.  Although it is difficult to write the distribution of
$\Ss$, we can approximate it with another sequence $\Ws$.  Let $B(k)$ be the
minimum of $k$ independent uniform $[0,1]$ random variables. Let $\Bs$ be a
sequence of i.i.d.\ random variables with distribution $B(k)$.  We define $W_t
= |S_0| \times \prod_{s = 1}^t B_s$.  

Given two random variables $A$ and $B$, we say $A$ is {\em stochastically
smaller than} $B$, denoted by $A \preceq B$, if and
only if
$$
\p{A \ge r} \le \p{B \ge r} \qquad \hbox{for all } r \in \R,
$$
where $\R$ is the set of real numbers.  The random variable $W_t$ is
stochastically larger than $|S_t|$, as there is a ``trimming'' effect at each
hop.  For example, the number of leaves between $z_1$ and $y'$ has a
distribution similar to $\lfloor W_1 \rfloor$. But some of these leaves might
not belong to $S_1$.

\begin{figure}
\centering {
    \scalebox{0.8} {
        \includegraphics{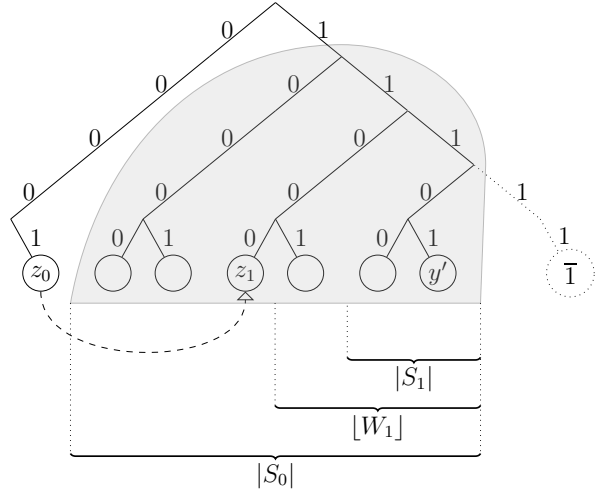}
    }
}
\caption{The ``trimming'' effect.}
\label{fig:cut}
\end{figure}

\begin{lemma} \label{lem:betabound}
For all $t \ge 1$, we have $|S_t| \preceq W_t$.
\end{lemma}

\begin{lemma} \label{lem:union:bound}
For all $t \in \N$, we have:
\begin{align*}
    & \hbox{(i)   } \qquad \sup_{x_1,\ldots,x_n} \sup_x \sup_y \p{T_{xy} \ge t} \le \p{n B_1 \ldots B_t \ge 1}, \cr
    & \hbox{(ii) }  \qquad \sup_{x_1,\ldots,x_n} \sup_x \p{\sup_y T_{xy} \ge t} \le n \times \p{n B_1 \ldots B_t \ge 1}, \cr
    & \hbox{(iii)}  \qquad \sup_{x_1,\ldots,x_n} \p{\sup_{x} \sup_y T_{xy} \ge t} \le n^2 \times \p{n B_1 \ldots B_t \ge 1}\enspace.
\end{align*}
\end{lemma}

A beta random variable $B(a,b)$ has probability density function
$$
f(x) = {{\Gamma(a+b)} \over {\Gamma(a)\Gamma(b)}}\,
x^{a-1}(1-x)^{b-1}, \quad 0 \le x \le 1,
$$
where $\Gamma(z)$ is the gamma function
$
\Gamma(z) = \int_0^\infty  e^{-t} t^{z-1} \, {\rm d}t.
$
In order statistics theory, a basic result~\cite[chap.2.3]{David2003} is that
the $r$-th smallest of $m$ i.i.d.\ uniform random variables has
beta distribution $B(r, m+1-r)$. Plugging in $r=1,m=k$, we have 
$
B_t \eql B(k) \eql B(1,k)
$
for all $t \in \N$. (For more about beta distribution,
see~\cite[chap.25]{Johnson1995}.) It is easy to check that for all $r > 0$ and
$t \in \N$, we have
\begin{equation}
\E{ \left( B_1 \cdots B_t \right)^r} 
= \E{B_1^r}^t = \E{B(1,k)^r}^t 
= \left( {{k!} \over {\prod_{i=1}^k (r+i)}} \right)^t \enspace.
    \label{eq:moment}
\end{equation}
By applying this moment bound, we have the following theorem:
\begin{theorem} \label{thm:speed}
There exist constants $\ck$,
$\ckp$ and $\cks$ such that: \hfill\break
(i) for all $c > \ck$,
$$
\sup_{x_1,\ldots,x_n} \sup_x \sup_y \p{T_{xy} \ge c \log n} \to 0 \qquad \hbox{as } n \to \infty;
$$
(ii) for all $c > \ckp$,
$$
\sup_{x_1,\ldots,x_n} \sup_x \p{\sup_{y} T_{xy} \ge c \log n} \to 0 \qquad \hbox{as } n \to \infty;
$$
(iii) for all $c > \cks$,
$$
\sup_{x_1,\ldots,x_n} \p{\sup_{x,y} T_{xy} \ge c \log n} \to 0 \qquad \hbox{as } n \to \infty\enspace.
$$
In particular, we have $\ck = 1/H_k$ where $H_k$ is the $k$-th harmonic number.
\end{theorem}

It is easy to check that $c_1' = e$, since ${(r+1) / {\log (1 + r)}}$ takes
minimum value $e$ when $r = e-1$. But unlike $\ck$, we do not have
closed forms for $\ckp$ and $\cks$. Table~\ref{table:alpha} shows the numerical values
of $\ck, \ckp, \cks$ for $k = 1,\ldots,10$.

\begin{table}
\caption{The numerical values of $\ck, \ckp, \cks$ for $k
    = 1,\ldots,10$.}
\label{table:alpha}
\setbox0=\vbox{
\halign{\hfil#\hfil&\quad#\hfil&\quad#\hfil&\quad#\hfil\cr
\tabletoprule
  $k$ & \hfill$\ck$\hfill & \hfill$\ckp$\hfill & \hfill$\cks$\hfill \cr
\tablerule
 1    & 1            & 2.718281828  & 3.591121477  \cr
 2    & 0.6666666667 & 1.673805050  & 2.170961287  \cr
 3    & 0.5454545455 & 1.302556173  & 1.668389781  \cr
 4    & 0.4800000000 & 1.105969343  & 1.403318015  \cr
 5    & 0.4379562044 & 0.9817977138 & 1.236481558  \cr
 6    & 0.4081632653 & 0.8950813294 & 1.120340102  \cr
 7    & 0.3856749311 & 0.8304602569 & 1.034040176  \cr
 8    & 0.3679369251 & 0.7800681679 & 0.9669189101 \cr
 9    & 0.3534857624 & 0.7394331755 & 0.9129238915 \cr
 10   & 0.3414171521 & 0.7058123636 & 0.8683482160 \cr
\tablebottomrule
}
}
\hbox to \hsize{\hfil\box0\hfil}
\end{table}

\begin{lemma} \label{lem:constants} We have
    $$
    \lim_{k \to \infty} \ck \log k = 
    \lim_{k \to \infty} \ckp \log k = 
    \lim_{k \to \infty} \cks \log k = 1\enspace.
    $$
\end{lemma}

\begin{remark} We are not providing precise inequalities with
matching lower bounds. This can be done, but in that case, one could have to
distinguish between many choices for $d$. We have already noted that $d \ge
\log_2 n$. We always have
$$
T_{xy} \le d\enspace.
$$
Therefore, for $d$ very large, there is a danger of having routing times that
are super-logarithmic in $n$. Our analysis shows that this is not the case.
However, the precise behavior of $T_{xy}$, uniformly over all $x$,$y$ and $d$,
requires additional analysis. The behavior for $d$ near $\log_2 n$,
$d=\Theta(\log n)$, and $d/\log n \to \infty$ is quite different.
\end{remark}

\begin{remark} The performance bounds of this section are of the form
$c_k \log n$ with $c_k = 1/H_k$. Although formulated for fixed $k$, they remain
valid if $k$ is allowed to depend upon $n$. For example, if $k = \log n$---that
is, the routing table size grows as $d \times \log n$---the expected routing
time is bounded by 
$$
(1+o(1)) { {\log n} \over {\log k}} \sim (1+o(1)){ {\log n} \over {\log \log n}}\enspace.
$$
Even more important is the possibility of having $O(1)$ routing time. With
$k \sim n^{\theta}$ for $\theta \in (0,1)$, the routing table size for one
computer grows as $d \times n^{\theta}$, and
$$
\E{\sup_x \sup_y T_{xy}} \le {1 \over \theta} + o(1)\enspace.
$$
The parameter $\theta$ can be tweaked to obtain an acceptable compromise
between storage and routing time.
\end{remark}

\section{The random ID model}
\label{sec:random}

In this section, we assume that $X_1,\ldots,X_n$ are chosen uniformly at random
from $\idspace$ without replacement. Recall that $\polar{X}_1$ denotes the \id\
that is farthest from $X_1$.  We sketch the proof that the distribution of
$\pT \eqd \pTi{1}$ has concentrated mass.

Since rotating the hypercube does not change the distribution of the routing
time, we can always assume that $X_1 = \origin$ and $\polar{X}_1 =
\polarorigin$, where $\bar{b}$ denotes the all-$b$ vector $(b,b,\ldots,b)$.

Write the routing path $\rho_{X_1\polar{X}_1} = (z_0,z_1,\ldots)$. Let $a_t$ be
the lowest common ancestor of $z_t$ and $y'$, i.e., the rightmost leaf in
the trie and the destination of routing. Then the sequence $(a_0,a_1,\ldots)$
can be seen as travelling downwards along the path from the root to $y'$, i.e.,
the rightmost branch of the trie, with the distance of each hop being random.

This sequence can be defined equivalently as follows. Let $a_0$ be the root of
the \id\ trie.  Let $z_0 = \origin$. From the right subtree of node $a_t$,
select up to $k$ paths to the bottom uniformly at random without replacement.
(This is equivalent to the choice of $k$ nodes to fill one \kbucket\ of $z_t$,
the $t$-th hop in the search.) If that right subtree is empty, then the search
terminates at $a_t$.  Let $z_{t+1}$ be the leaf corresponding to the rightmost
one of these selected paths. Let $a_{t+1}$ be the lowest common ancestor of
$z_{t+1}$ and $y'$.  Let $L_{t+1}$ be the distance from $a_0$ to $a_{t+1}$. Let
$R_{t+1} = L_{t+1} - L_{t}$, i.e., the distance of the $(t+1)$-th hop, which is
also the distance between $a_t$ and $a_{t+1}$.  Note that $\pT = t$ if and only
if $\sum_{i=1}^t R_i=d$. Therefore, we can bound $\pT$ by studying the
properties of $\seqone{R_t}$.

Now instead of the \id\ trie of depth $d$, consider a full binary trie with
infinite depth. Definite $(a_0',a_1',\ldots)$, the counterpart of
$(a_0,a_1,\ldots)$ in this infinite trie, as follows.  Let $a_0'$ be the root
of the infinite trie. From the right subtree of node $a_t'$, select exactly $k$
infinite downwards paths uniformly at random.  (Since the probability of
selecting the same path more than once is zero, ``without replacement'' is not
necessary anymore.) Let $z_{t+1}'$ be the rightmost of these selected paths.
Let $a_{t+1}'$ be the lowest common node of $z_{t+1}'$ and $\polarorigin$.  Let
$L_{t+1}'$ be the distance from $a_0'$ to $a_{t+1}'$. Let $G_{t+1} = L_{t+1}' -
L_{t}'$, i.e., the distance of $(t+1)$-th hop, which is also the distance
between $a_t'$ and $a_{t+1}'$.

When $n$ is large (and thus $d$ is large), the behavior of $\seqone{R_t}$ and
$\seqone{G_t}$ are very similar.  But it is easy to see that, since each
subtree of the infinite trie has exactly the same structure, $\seqone{G_t}$ is a
sequence of i.i.d.\ random variables with distribution
\begin{equation}
\p{G_1 \le i} = \left( 1 - {1 \over {2^i}} \right)^k \qquad \hbox{for all } i \in
\N.
    \label{eq:g:dist}
\end{equation}
In other words, $\seqone{G_t}$ is much easier to analyze. (Note that when
$k=1$, $G_1$ is simply the geometric distribution.) And it is possible to
couple the random variables $G_t$ and $R_t$.

When the downwards travel reaches the depth of $\log_2 n$, the routing can not
last much longer. In fact, in the \id\ trie, a subtree whose root has depth at
least $\log_2
n$ has only $o(\log n)$ leaves with high probability \cite{Cai2012}.
Therefore, we define
$$
T_n = \min \left\{t : t \ge 1, \sum_{i=1}^t G_i \ge \log_2 n \right\}.
$$
\begin{lemma} \label{lem:t:n} We have
    $$
    \E{ {T_n} \over {\log n}} \to {1 \over {\log(2) \times \E{G_1}}}, 
    $$
    as $n \to \infty$, and also
    $$
    { {T_n} \over {\log n}} \to {1 \over {\log(2) \times \E{G_1}}} \qquad
    \hbox{in probability}, 
    $$
    as $n \to \infty$.
\end{lemma}
\noindent Then, by coupling, we can show Lemma~\ref{lem:converge}:
\begin{lemma} \label{lem:converge}
    We have
    $$
    {{\pT - T_n} \over {\log n}} \to 0 \qquad \hbox{in probability},
    $$
    as $n \to \infty$.
\end{lemma}
\noindent We omit the proof of these two lemmas due to space limitations. For details, see
\cite{Cai2012}. (For other proofs, see the appendix.)

\bibliography{citation}{}
\bibliographystyle{splncs}

\newpage

\centerline{\twelvebf Appendix}

\subsubsection*{Proof of Lemma~\ref{lem:betabound}}

\begin{proof}
Let $m,b$ be two positive integers.
Let $Z(m,b)$ be the minimum of a sample of size up to $m$ selected
uniformly at random from $\{0,1,\ldots,b-1\}$ without replacement.
It is easy to check that
$$
Z(m,b) \preceq b \times B(m) \qquad \hbox{for all } m,b \in \N,
$$
where $\N$ denotes the set of positive integers.

Consider step $t$ in the routing algorithm.  Let $Z_t$ be the number of leaves
to the right of $z_t$. Since up to $k$ leaves are selected uniformly at random
without replacement from $S_{t-1}$ and the rightmost one is chosen as $z_t$,
$Z_t$ has distribution $Z(k,|S_{t-1}|)$.  Since $Z_t \ge |S_t|$, it follows
that for all $r \in \R$ and $b \in \N$, we have
\begin{align*}
\p{|S_t| \ge r~|~|S_{t-1}| = b}
& \le \p{Z_t \ge r ~|~|S_{t-1}| = b}  \cr
& \le \p{b \times B(k) \ge r} \cr
& = \p{b \times B_t \ge r}\enspace.
\end{align*}
Summing over all possible $b$, we have
\begin{align*}
    \p{ | S_t | \ge r } 
        &= \sum_{b \in \N} 
        \p{ | S_t | \ge r ~|~ | S_{t-1} | = b } 
        \p{ | S_{t-1} | = b} \cr
        &\le \sum_{b \in \N} \p{ b \times B_t \ge r } \p{ | S_{t-1} | = b} \cr
        &= \sum_{b \in \N} \p{ b  \times B_t \ge  r \cap | S_{t-1} | = b} \cr
        &= \p{ B_t \times | S_{t-1} | \ge r},
\end{align*}
which implies $|S_t| \preceq B_t \times |S_{t-1}|$.

To finish the proof, we need a simple result from stochastic order theory~\cite[thm.1.A.3]{Shaked2007}.
Given two nonnegative random variables $Y_1$ and
$Y_2$ with $Y_1 \preceq Y_2$, and a random variable $Z$ independent of both $Y_1$
and $Y_2$, we have
$$
Y_1 \times Z \preceq Y_2 \times Z\enspace. 
$$
Recursively applying this inequality, we have
$$
| S_t | 
\, \preceq \, B_t | S_{t-1} |
\, \preceq \, B_t B_{t-1} | S_{t-2} |
\, \preceq \, \dots
\, \preceq \, \prod_{s=1}^t B_s |S_0|
\, =  \, W_t\enspace. \quad \square
$$
\end{proof}

\subsubsection*{Proof of Lemma~\ref{lem:union:bound}}

\begin{proof} Note that $T_{xy} \ge
t$ if and only if $|S_t| \ge 1$. It follows from the previous lemma that
$$
\p{T_{xy} \ge t} = \p{|S_t| \ge 1} \le \p{|S_0| B_1 \cdots B_t \ge 1} \le \p{n
B_1 \cdots B_t \ge 1}\enspace.
$$
Clearly (ii) and (iii) follow from the union bound. $\square$
\end{proof}

\subsubsection*{Proof of Lemma~\ref{lem:constants}}

\begin{proof} For $c_k$, this follows from $\ck = 1/H_k$. For $\ckp$ and
$\cks$, note first
$$
\ck \le \ckp \le \cks\enspace.
$$
Next in the definition of $\ckp$ and $\cks$, we obtain upper bounds by
specifying a certain value of for $r$. So,
$$
{1 \over {H_K}} \le \ckp \le \cks 
\le { {\sqrt{\log k}+2} \over \sum_{i=1}^k \log(1 + \sqrt{\log k} / i)}
= (1+o(1)) {1 \over {\log k}}.
$$
The last equality follows easily by noticing that for all $\epsilon > 0$, there
exists $x_{0} > 0$, such that
$$
\log(1+x) \ge (1-\epsilon)x
$$
for $0 \le x \le x_0$. Thus, provided $k$ is so large that ${1 \over
    {\sqrt{\log k}}} \le x_0$, we have
\begin{align*}
    \sum_{i=1}^k \log \left( 1 + { {\sqrt{\log k}} \over i }\right)
    & \ge \sum_{i= \lceil \log k \rceil}^k 
    \log \left( 1 + { {\sqrt{\log k}} \over i }\right) \cr
    & \ge (1-\epsilon) \sum_{i= \lceil \log k \rceil}^k 
    { {\sqrt{\log k}} \over i } \cr
    & = (1-\epsilon)(H_{k}-H_{\lceil \log k \rceil})\sqrt{\log k} \cr
    & \sim (1-\epsilon)(\log k)^{3/2}.
\end{align*}
Therefore,
$$
\cks \le { {\sqrt{\log k} + 2} \over {(1-\epsilon)}(\log k)^{3/2}(1+o(1))} 
= { {1+o(1)} \over {(1-\epsilon) \log k} }.
$$
Since $\epsilon$ is arbitrary, we are done. $\square$
\end{proof}

\subsubsection*{Proof of Theorem~\ref{thm:speed}}

\begin{proof}
It follows from (i) of Lemma~\ref{lem:union:bound}, the moment bound and equation~\eqref{eq:moment} that for all
and $r > 0$ and $t \in N$, we have
\begin{equation}
\p{T_{xy} \ge t} 
\le \p{n B_1 \cdots B_t \ge 1} 
\le \E{B_1^r} n^r 
= \left( {{k!} \over {\prod_{i=1}^k (r+i)}} \right)^t n^r\enspace.
    \label{eq:moment:bound}
\end{equation}
Taking $t = c \log n$, we have
$$
\log \left( \p{T_{xy} \ge c \log n} \right) \le 
\left(r - {{c} \times  {\sum_{i=1}^k \log(1+r/i)}} \right) \log n\enspace.
$$
Therefore, $\p{T_{xy} \ge c \log n} \to 0$ as $n \to \infty$ if
$$
r - {{c} \times  {\sum_{i=1}^k \log(1+r/i)}} < 0\enspace.
$$
Solving this inequality, we see that we need
$$
c >  h_k(r) \eqd {r \over {\sum_{i=1}^k \log (1 + r/i)}}\enspace.
$$
Thus, (i) follows if we take $\ck = \inf_{r > 0} h_k(r)$. Simple
calculations show that $h_k(r)$ increases on $(0,\infty)$. Therefore,
$$
\ck =  \lim_{r \downarrow 0} h_k(r) =  \lim_{r \downarrow 0} {1 \over {\sum_{i=1}^k\log (1+r/i)^{1/r}}} =
{1 \over {\sum_{i=1}^k 1/i}} = {1 \over {H_k}}\enspace.
$$
By a similar argument, (ii) and (iii) follows if we take
$$
\ckp = \inf_{r > 0} {{r+1} \over {\sum_{i=1}^k \log (1 + r/i)}}, \qquad
\cks = \inf_{r > 0} {{r+2} \over {\sum_{i=1}^k \log (1 + r/i)}}\enspace. \quad \square
$$
\end{proof}

\subsubsection*{Proof of Theorem~\ref{thm:speed:upper}}

\begin{proof} For all $x \in \Vertices$, $y \in
\idspace$ and $c > \ck$, we have
\begin{align*}
\E{T_{xy}} 
& = \sum_{t = 1}^\infty \p{T_{xy} \ge t} \cr
& = \sum_{t = 1}^{\lfloor c \log n \rfloor} \p{T_{xy} \ge t}  
+ \sum_{t = \lceil c \log n \rceil}^\infty \p{T_{xy} \ge t} \cr
& \le c \log n + \sum_{t = \lceil c \log n \rceil}^\infty \left( {{k!} \over {\prod_{i=1}^k (r+i)}} \right)^t n^r,
\qquad \hbox{for all } (r > 0).
\end{align*}
The last inequality follows from the fact that probability is at most $1$
and equation~\eqref{eq:moment:bound}. Note that the second term is a
geometric series. We already showed that the first term in the series
tends to $0$ as $n \to \infty$. The ratio between consecutive terms is
$k!/\prod_{i=1}^k (r+i) < 1$ for every fixed $r > 0, k \ge 1$.
Thus for all $\epsilon > 0$,
there exist $n_0 \in \N$, such that for all $n > n_0$ we have
$$
\E{T_{xy}} \le (\ck + \epsilon) \log n,
$$
or simply
$$
\E{T_{xy}} \le (\ck + o(1)) \log n\enspace.
$$
(ii) and (iii) follow by similar arguments. $\square$
\end{proof}

\subsubsection*{Proof of Theorem~\ref{thm:distribution:t:polar}}

\begin{proof}
It follows from \eqref{eq:g:dist} that
$$
\E{G_1} \ge \int_0^\infty 1-\left(1- {1 \over {2^x} } \right)^k {\rm d}x 
= { {H_k} \over {\log 2}},
$$
and
$$
\E{G_1} \le 1 + \int_0^\infty 1-\left(1- {1 \over {2^x} } \right)^k {\rm d}x 
= 1 + { {H_k} \over {\log 2}}\enspace.
$$
Therefore,
$$
g(k) \eqd \log(2) \times \E{G_1} = H_k + O(1)\enspace.
$$
It follows from Lemma~\ref{lem:t:n} that Lemma~\ref{lem:converge} that
$$
{ {\pT} \over {\log n}} \to {1 \over g(k)}
    \qquad \hbox{in probablity},
$$
as $n \to \infty$. $\square$
\end{proof}

\end{document}